\documentclass[10pt,preprint2]{aastex}

\def\figdir{.}

\def\ion#1#2{\mbox{\rm #1\sc #2}}

\def\HII{{\ion{H}{ii}}}

\def\dim#1{\mbox{\,#1}}

\def\figname#1{\figdir/#1}

\begin{document}

\title{Reionization, SLOAN, and WMAP: is the Picture Consistent?}
\author{Nickolay Y.\ Gnedin}
\affil{Center for Astrophysics and Space Astronomy, 
University of Colorado, Boulder, CO 80309}
\email{gnedin@casa.colorado.edu}

\def\dim#1{\mbox{\,#1}}

\label{firstpage}

\begin{abstract}
I show that advanced simulations of cosmological reionization are able to
fit the observed data on the mean transmitted flux in the hydrogen
Lyman-alpha line at $z\sim6$. At the same time, posteriori models can be
constructed that also produce a large value (20\%) for the Thompson
scattering optical depth, consistent with the WMAP measurements. Thus, it
appears that a consistent picture emerges in which early reionization (as
suggested by WMAP) is complete by $z\sim6$ in accord with the SLOAN data.
\end{abstract}


\section{Introduction}

Are we knocking at the door of reionization? At least, it seems so, as the
SLOAN quasar survey continues to find ever more distant quasars (Becker et
al.\ 2001; Fan et a.\ 2002, 2003; White et al.\ 2003). In fact, 
highest redshift quasars show essentially no transmitted flux just blueward 
of the quasar \HII\ region, which has been interpreted as the evidence that
that reionization took place at a redshift $z\sim6$ (Becker et
al.\ 2001; White et al.\ 2003; a similar claim was also made by Djorgovski
et al.\ 2001 based on an extrapolation from a lower redshift observations).

However, one should be cautious before drawing such a conclusion. Indeed,
the observed decrease in the mean transmitted flux at $z\sim 6$ might
simply indicate a decrease in the mean ionizing intensity rather then a real
reionization of the universe. After all, a neutral fraction of only
$2\times10^{-4}$ is sufficient to absorb effectively all Lyman-alpha
radiation at $z=6$ even in the gas that is underdense by a factor of 10.
Thus, without an understanding of the
evolution of the universe around the reionization epoch, the Ly-alpha
absorption data cannot be used to constrain the epoch of reionization
(unless damping wings are observed in the absorption profiles, 
Miralda-Escud\'{e} 1998).

Fortunately, our theoretical understanding of the process of reionization,
in part based on numerical simulations, is solid enough so that the
evolution of the ionizing intensity at these redshifts can be predicted
with a reasonable confidence level. Combining simulations with the
observational data indeed allows one to come up with a consistent picture
for the process of cosmological reionization.

\section{Simulations}

Four sets of simulations have been performed with
the SLH code and are similar to the simulations reported in
Gnedin (2000). The main difference with previous simulations is that
a newly developed and highly accurate
Optically Thin Eddington Variable Tensor (OTVET) 
approximation for modeling radiative transfer (Gnedin \& Abel 2001)
is used instead
of a crude Local Optical Depth approximation. The new simulations therefore
should be sufficiently accurate (subject to the 
usual limitations of numerical convergence and phenomenological description
of star formation) to be used meaningfully in comparing with the
observational data. In particular, my fiducial model used the cosmological
parameters as determined by the {\it WMAP} satellite (Spergel, et al. 2003).

\begin{table}
\caption{Simulation Parameters\label{sim}}
\begin{tabular}{lcccc}
\tableline
Set & 
$\Omega_{m}$ & 
$h$ &
$n$ & 
Box size ($h^{-1}\dim{Mpc}$) \\
\tableline
\tableline
A4 & 0.27 & 0.71 & 1.0  & 4 \\
A8 & 0.27 & 0.71 & 1.0  & 8 \\
B4 & 0.35 & 0.70 & 0.95 & 4 \\
C4 & 0.35 & 0.70 & 0.97 & 4 \\
\tableline
\end{tabular}
\end{table}
Parameters of the four sets are given in Table \ref{sim}.
All simulations included $128^3$ dark matter particles, an equal
number of baryonic cells on a quasi-Lagrangian moving mesh, and about
3 million stellar particles that formed continuously during the simulation.
The nominal
spatial resolution of simulations with the box size of $4h^{-1}\dim{Mpc}$
 was fixed at $1h^{-1}$ comoving kpc, with the real resolution being a
 factor of two worse. Simulations with the box size of $8h^{-1}\dim{Mpc}$
 had the twice worse spatial resolution.

In all cases a flat cosmology was assumed, with $\Omega_{\Lambda,0} =
1-\Omega_{m,0}$, and normalization of the primordial fluctuations was
determined either from the {\it WMAP} data (Spergel et al.\ 2003) for sets A4 and
A8, or from the {\it COBE} data (White \& Bunn 1995).
Notice that a small change in the slope of the primordial power-law
spectrum $n$ makes a significant effect on the amount of the small-scale
power due to a large leverage arm from CMB scales to the tens-of-kpc
scales which are important for reionization.

Star formation is incorporated in the simulations using a
phenomenological Schmidt law, which introduces two free parameters:
the star formation efficiency $\epsilon_{\rm SF}$ (as defined by eq.\ (1)
of Gnedin 2000) and the ionizing radiation efficiency
$\epsilon_{\rm UV}$ (defined as the energy in ionizing photons per unit of
the rest energy of stellar particles).

The star formation efficiency $\epsilon_{\rm SF}$ 
is chosen so as to normalize the global star
formation rate in the simulation at $z=4$ to the observed value from
Steidel et al.\ (2001), whereas the ultraviolet radiation
efficiency $\epsilon_{\rm UV}$
is only weakly constrained by the (highly uncertain) mean
photoionization rate at $z\sim4$. The redshift of reionization
strongly depends on $\epsilon_{\rm UV}$ and is not in fact predicted in a
simulation, but can be changed over a reasonable range depending on the
assumed value of $\epsilon_{\rm UV}$. Each of the simulations sets included
in Table \ref{sim} in fact included several individual simulations with
different values of $\epsilon_{\rm UV}$.

However, since the simulations are quite expensive, it is not possible to
cover a large range of $\epsilon_{\rm UV}$ in a given set. Typically, only
2 or 3 simulations per set have been performed and the results are then
interpolated between the simulations. This procedure is fully described in
\S \ref{results}.

\section{A ``Redshift of Reionization'': What Is It?}

\begin{figure}[t]
\plotone{\figname{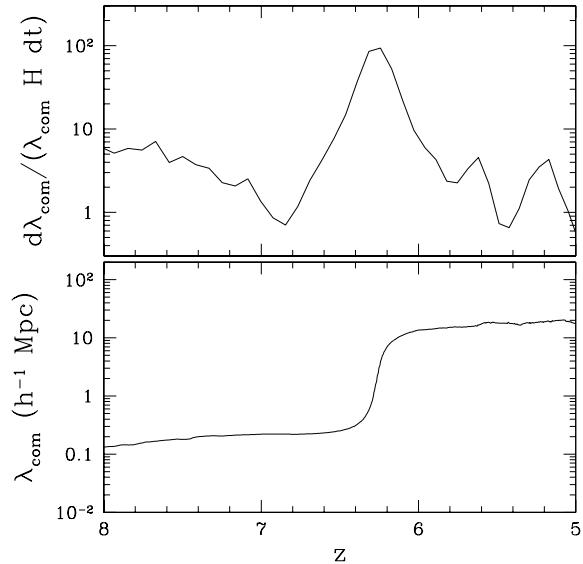}}
\caption{Mean free path to ionizing radiation (bottom) and its time derivative
(top) as a function of redshift for the simulation A4 from
Table \protect\ref{sim}.} 
\label{figMP}
\end{figure}
Reionization is a process, not an event. In fact, whole process of
reionization is quite extended ($\Delta z\sim5-10$), and can be generically
separated into three stages:
(i) the {\bf pre-overlap stage}, in which individual \HII\ regions around the
  sources of ionization expand and merge in the low density IGM,
(ii) the {\bf overlap stage}, in which all individual \HII\ regions
  overlap, and 
  last remnants of the neutral low density gas quickly disappear, and
(iii) the {\bf post-overlap stage}, in which remaining high density gas is
  being 
  ionized from the outside, until neutral gas remains only in some of the
  highest density regions, which would be identified as Lyman-limit system
  in the absorption spectra of distant quasars. 

One can think reionization as complete, when the mean free path to the
ionizing radiation is fully determined by (relatively) 
slowly evolving Lyman-limit systems (Miralda-Escud\'e 2003a).

It is, however, tempting to try to assign a value for the ``redshift of
reionization''. Such a value, in order not to be completely arbitrary, must
be related to the physics of the reionization process. For example, it
seems to be natural to identify the moment of reionization with the overlap
stage, but even it takes place over a sizable redshift interval 
$\Delta z\sim1$ (Gnedin 2000), so we would need a better definition if we
are to assign a value to ``the redshift of reionization''.

Fortunately, such a definition exists. Figure \ref{figMP} shows the mean
free path to ionizing radiation and its time derivative as a function of
redshift for the fiducial simulation (A4) with $\epsilon_{\rm
  UV}=1\times10^{-6}$. The time derivative of the mean free path has a
well-defined peak in the middle of the overlap stage, which is a natural
moment to identify with ``the redshift of reionization'' $z_{\rm REI}$. 

\begin{figure}[t]
\plotone{\figname{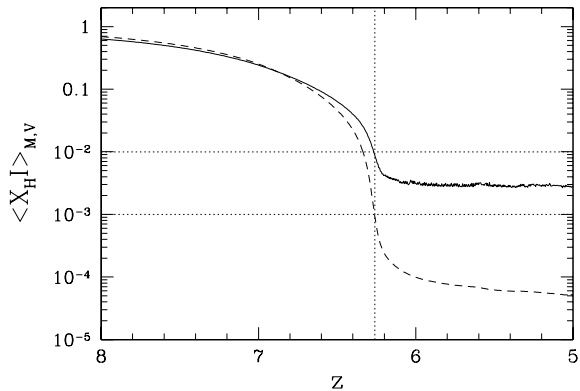}}
\caption{Mass (solid line) and volume (dashed line) weighted neutral
  hydrogen fraction as a function of redshift 
for the simulation A4 from Table \protect\ref{sim}. Notice that the peak in
  the time derivative of the mean free path (vertical dashed line) closely
  corresponds to the moment when the mean mass (volume) weighted neutral
  fraction  reaches a value of $10^{-2}$ ($10^{-3}$) respectively.}
\label{figXH}
\end{figure}
However, this definition is not entirely practical, since the time
derivative of the mean free path cannot be observed directly. Instead, a
more easily observable quantity is the mean neutral fraction. Figure
\ref{figXH} shows the mean mass and volume weighted neutral hydrogen
fraction for the fiducial simulation. The moment of reionization closely
corresponds to the time when the mean mass (volume) weighted neutral
fraction reaches a value of $10^{-2}$ ($10^{-3}$) respectively - one can
consider that moment as an alternative definition to ``the redshift of
reionization''. 

The latter definition is more practical, but it is a subject to an
important clause: while the volume weighted neutral fraction is reliably
computed in the simulation, the mass weighted one depends on the numerical
resolution. In particular, the simulations presented in this paper do not
resolve the damped Lyman-alpha systems, so the quoted above value of 1\%
for the mass weighted neutral fraction does not include the neutral gas
locked in the damped Lyman-alpha systems. In this respect, the volume
weighted number is more robust and should be used as a main definition of
``the redshift of reionization''.

\section{Results}
\label{results}

\subsection{Fitting the Mean Transmitted Flux Data}

\begin{figure}[t]
\plotone{\figname{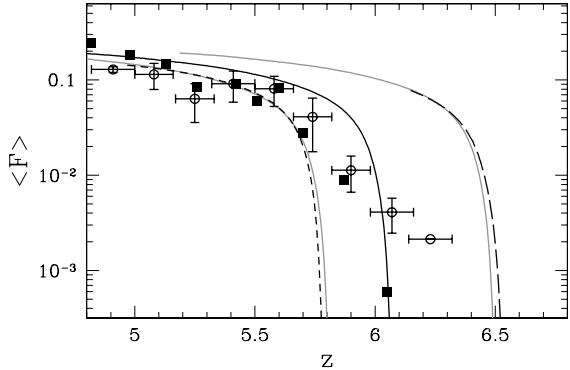}}
\caption{Mean transmitted flux as a function of redshift
for simulations from the set B4. Solid, long-dashed, and short-dashed black
lines correspond to simulations with $\epsilon_{\rm UV}=2.5\times10^{-6}$,
$4.0\times10^{-6}$, and $2.0\times10^{-6}$ respectively. Two gray lines are
the solid black line simply shifted horizontally to match other two lines.
Open circles with error-bars show the observational data from White et al.\
(2003) and filled squares are the data of Songaila (2004; the error-bars
are not shown for these points for clarity).}
\label{figTZ}
\end{figure}
Figure \ref{figTZ} shows the mean transmitted flux as a function of
redshift for the three simulations with different values of the
$\epsilon_{\rm UV}$ parameter from the set B4. The observational data
shows with open circles were obtained from White et al.\ (2003) by
averaging the mean transmitted flux at a given redshift interval. The
vertical error-bars are errors of the mean (not mean errors!). The
horizontal error-bars are simply the width of the redshift interval over
which the mean transmitted flux is computed. The last data point (without
the vertical error-bar) is likely to be contaminated by a foreground galaxy
and is not included in this analysis. Filled squares show the data from
Songaila (2004). In the latter case I do not show the error-bars for
clarity, but they are comparable to that of White et al.\ (2003).

Two gray lines are the solid
black line simply shifted horizontally. As one can see, the change in
the $\epsilon_{\rm UV}$ parameter simply translates into the shift in
redshift. This is not surprising given that in the $\Lambda$CDM model
clustering proceeds hierarchically, in a quasi self similar manner. In
other words, one would expect that the dependence of the mean transmitted
flux on the emissivity parameter $\epsilon_{\rm UV}$ and redshift $z$ enters,
to the first order, only as 
\begin{equation}
\langle F\rangle(z) = (1+z)^\beta g\left[\epsilon_{\rm UV}(1+z)^{-\alpha}
\right],
\label{epsuvvsz}
\end{equation}
where $g$ is a function of one argument, and 
$\alpha$ and $\beta$ depend on the slope of the
primordial power spectrum at the scale of interest, and may also depend on
the details of the temperature-density relation. But because I am only
concerned with a narrow redshift range around $z=6$ (for example,
$z=5.8\div6.5$ in Fig.\ \ref{figTZ}), $\alpha$ and $\beta$ from equation
(\ref{epsuvvsz})
can be considered constant in the first approximation. Fig.\ \ref{figTZ}
indicates that 
for the range of redshifts considered $\beta$ is small, less than
about 0.5 in the absolute value. The specific reason for such a small value
is not entirely clear, and may be a simple coincidence: if $\epsilon_{\rm
  UV}$ is reduced, the amount of ionizing radiation in the post-overlap
stage is also reduced, but at the same time the post-overlap stage is
delayed to lower redshifts, when densities are lower. The two effects
appear to approximately cancel each other, leading to a small value of
$\beta$. 

The main result of Fig.\ \ref{figTZ} is that a change in the emissivity
parameter 
$\epsilon_{\rm UV}$ by a factor of $f$ simply translates into a rescale of
(one plus) redshift by a factor $f^{1/\alpha}$.
Note that such rescaling does not affect the vertical amplitude of the mean
transmitted flux curve. As is well known, the mean absorption after the
post-overlap stage is dominated by the Lyman-limit systems
(Miralda-Escud\'e 2003b). Thus, in order to reproduce the observed run of the
mean transmitted flux with redshift, a simulation must have (i) the right
value for the emissivity parameter $\epsilon_{\rm UV}$ (i.e.\, the
right value for the redshift of reionization), to reproduce the drop-off 
at $z\sim 6$, and (ii) the correct abundance of the Lyman-limit systems. 
The latter is crucially dependent on the abundance of low mass virialized
objects, and is {\it not\/} completely controlled by the parameter
$\epsilon_{\rm UV}$, but also depends
on the mass function of virialized objects, i.e.\ on cosmological
parameters. Therefore, if a simulation has wrong values of cosmological
parameters, it might not be able to reproduce the amplitude of the mean
transmitted flux as a function of redshift for {\it any} value of the
single free parameter $\epsilon_{\rm UV}$.

\begin{figure}[t]
\plotone{\figname{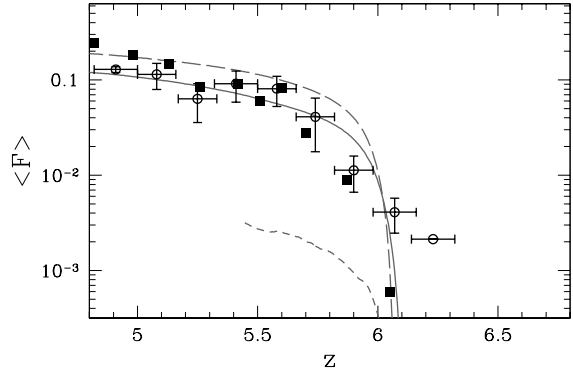}}
\caption{Mean transmitted flux as a function of redshift
for best-fit simulations from sets A4 (solid line), B4 (long-dashed
line), and C4 (short-dashed line). }
\label{figTR}
\end{figure}
This is illustrated in Figure \ref{figTR} which shows the results from
three simulation 
sets (A4, B4, and C4), each rescaled for the best value of the emissivity
parameter $\epsilon_{\rm UV}$ (the simulation set C4 has not been continued
beyond $z=5.5$). As one can see, not all of the simulation
sets are able to reproduce the amplitude of the mean transmitted flux curve.
For example, the set C4 predicts a factor of 10 more
absorption: in that cosmological model ($\Omega_m=0.35$, $n=0.97$) the
amount of small-scale power is significantly larger than, say, in the {\it
  WMAP} model, and the mean distance between Lyman-limit systems is a
factor of 10 less than the observed value. Sets A4 and B4, on the other
hand, provide a reasonable, although not perfect, fit to the data. In
fact, it appears that the set A4 (the WMAP model) fits the data best for
$z>5.2$, but goes somewhat below the data points at lower redshifts. This
is expected, since due to the limited size of the computational box,
simulations should unpredict the abundance of ionizing sources at lower
redshifts. The simulation set B4 has a somewhat higher mean transmitted
flux, although it is marginally consistent with the data given the cosmic
variance. 

There is no easy way to reconcile model C4 (and, perhaps, model B4) 
with the data. The
abundance of the Lyman-limit systems depends not only on $\epsilon_{\rm
  UV}$, but also on the cosmological model, and, as long as $\epsilon_{\rm
  UV}$ and cosmological parameters are specified, there is no free
parameter left to adjust this abundance. The only way to make, say, model
B4 to fit the data better 
would be to include additional Lyman-limit absorption
posteriori, by hand - for example, by postulating an entirely different
class of objects that contribute to the mean free path but do not form
within the hierarchical clustering framework. For model C4, however, one
would have to postulate that some of the Lyman-limit system that form in
the simulation do not form in reality.

\subsection{Measuring the Redshift of Reionization}

\begin{figure}[t]
\plotone{\figname{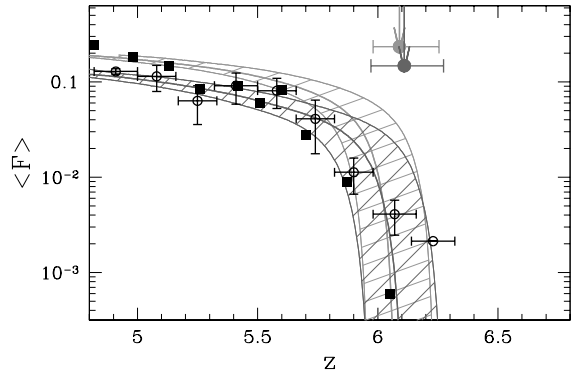}}
\caption{Mean transmitted flux as a function of redshift
for the best-fit simulations from sets A4 (dark gray) and B4 (medium gray).
Shaded regions illustrate the uncertainty due to observational
errors. Arrows with error-bars give the redshift of reionization plus its
$1\sigma$ uncertainty for each of the simulation sets.}
\label{figTA}
\end{figure}
Using the fact that the WMAP model fits the observed data, one can
determine how much the redshift of reionization can be shifted while still
preserving the acceptable fit to the data. Figure \ref{figTA} shows the
range of values for the redshift of reionization for models A4 and B4
(although the latter is admittedly a worse fit to the data) 
obtained by adding
linearly two different components: (i) an
average observational error-bar in $z$-direction (0.085), and
(ii) the average uncertainty to the best
fit value of $z_{\rm REI}$ for the fiducial model. With these three
uncertainties included, a measurement of the redshift of 
reionization can be made:
\begin{equation}
z_{\rm REI} = 6.1 \pm 0.3\ \ \mbox{(95\%~CL)}.
\label{res}
\end{equation}
If other cosmological models are also considered (for example, model B4),
then an additional ``systematic'' uncertainty is introduced due to small
differences in the redshifts of reionization of different cosmological
models. This uncertainty, however, is rather small, as can be seen from
Fig.\ \ref{figTA}, and is ignored here.

\subsection{Sanity Checks}

\begin{figure}[t]
\plotone{\figname{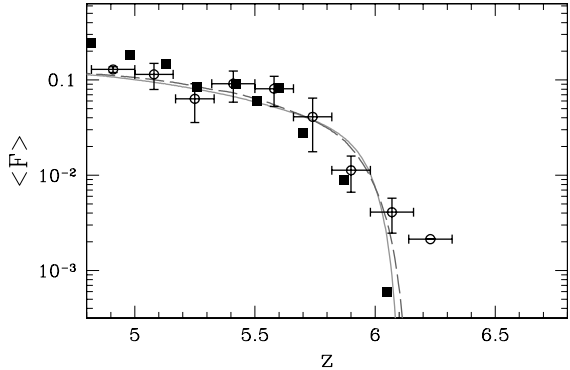}}
\caption{Tests of numerical convergence and cosmic variance.
Light gray line with error-bars is the best fit A4 set with the expected
rms fluctuation in the mean transmitted flux over a redshift interval
$\Delta z=0.17$ (which is the the average redshift interval over which the
transmitted flux is averaged in the observational data). 
The dark gray long-dashed line is the best-fit mean
transmitted flux for the A8 set (twice larger simulation volume).}
\label{figTC}
\end{figure}
Before I can proceed further, several sanity tests must be completed. 
The first two are shown in Figure
\ref{figTC}. The light gray line is again the best-fit fiducial {\it WMAP}
model, with error-bars showing the rms fluctuation in the mean transmitted
flux (error of the mean) - this is consistent with (although slightly
smaller than) the observed values. This is not surprising, as a finite (and
relatively small) size of the computational volume will lead to an
underestimate of the cosmic variance. 

How significant this overestimate? In order to reduce the
cosmic variance at the box size scale, the specific realizations of initial
conditions are selected to reproduce the correct power spectrum of initial
perturbations on the fundamental mode (Gnedin \& Hamilton 2002), which
mitigates the uncertainty due to cosmic variance by about a factor of 10 or
more. To demonstrate that numerical convergence is achieved for the mean
transmitted flux, and that cosmic variance does not affect my results, I
show with the dark gray line in Fig.\ \ref{figTC} the best-fit simulation
from the 
A8 set (the same cosmological model, twice larger box). The fact that the
two simulations with the two box sizes are sufficiently close argues for
the numerical convergence of the simulation result, although, admittedly,
this is not a
100\% rigorous test (three different resolutions would be required for a
complete test, but this is currently beyond the existing computer
capabilities). 

\begin{figure}[t]
\plotone{\figname{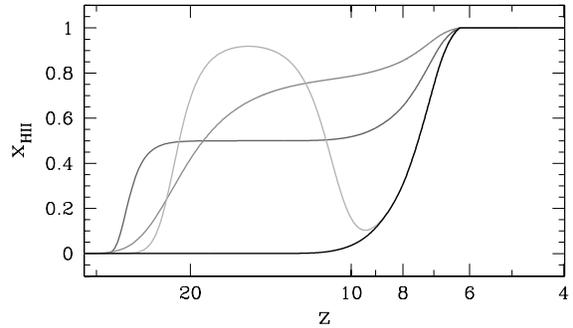}}
\caption{Ionization histories (volume averaged ionized fraction as a
function of redshift) for the fiducial model (black line), and three
arbitrary ionization histories (grey lines) with $\tau=0.2$. All four model
have the identical ionization redshift of $z_{\rm REI}=6.1$.}
\label{figXW}
\end{figure}
The recent results from the WMAP measurement of the cross-correlation
between the temperature and polarization anisotropies of the CMB indicate
the large value of the Compton optical depth $\tau=0.12\pm0.06$ (Kogut et
al.\ 2003, Tegmark et al.\ 2004). Does this observation rule out
reionization at $z\sim6$? Figure \ref{figXW} shows the ionization histories
for the fiducial model (which has $\tau=0.06$) together with three other
posteriori ionization histories which all have $\tau=0.2$. Because these
histories are adjusted to match the fiducial model before the volume
averaged ionized fraction approaches 99.9\%, they all have identical
redshifts of reionization and identical mean transmitted flux curves at
$z<6.1$. Thus, the WMAP measurement by itself neither constraints the
redshift of reionization nor contradicts the SLOAN data, but rather
indicates that the early ionization history might have been quite complex.

\section{Conclusions}

I have shown that advanced self-consistent cosmological simulation with
radiative transfer are able to reproduce the observed data on the evolution
of the mean transmitted flux from the SLOAN survey when the values of
cosmological parameters that best fit the WMAP data are adopted for the
underlying cosmology. Thus, a consistent picture is emerging in which the
universe reionizes at $6.1\pm0.3$ after possibly a prolonged period of
partial reionization.

\acknowledgements
I am grateful to the anonymous referee for correcting errors in the
original manuscript, and for other valuable comments.
This work was supported by NSF grant AST-0134373 and by 
National Computational
Science Alliance under grant AST-020018N and utilized 
SGI Origin 2000 array and IBM P690 array
at the National Center for Supercomputing Applications.

\end{document}